\def\rhodd{\rho_\mathrm{DDISC}}
\def\rhoh{\rho_\mathrm{HALO}}

\def\MWone{MW1}
\def\MWtwo{H204}
\def\MWthree{H258}
\def\MWthreedark{H258dark}

\def\vmaxd{v_\mathrm{max}}

\def\ltsima{$\; \buildrel < \over \sim \;$}
\def\simlt{\lower.5ex\hbox{\ltsima}}   
\def\gtsima{$\; \buildrel > \over \sim \;$}
\def\simgt{\lower.5ex\hbox{\gtsima}}

\newlength{\gsize}

\documentclass[cits]{PoS}
\title{A dark matter disc in the Milky Way}

\usepackage{myaasmacros}
\usepackage{subfigure}

\ShortTitle{A dark matter disc in the Milky Way}
\author{\speaker{J. I. Read}$^1$, V. Debattista$^3$, O. Agertz$^1$, L. Mayer$^{1}$, A. M. Brooks$^4$, F. Governato$^3$ and G. Lake$^1$\\
	$^1$Institute for Theoretical Physics, University of Zurich, Winterthurerstrasse 190 8047\\
	$^2$RCUK Fellow; Centre For Astrophysics, University of Central Lancashire, Preston, PR1 2HE\\
	$^3$Astronomy Dept., University of Washington, Box 351580, Seattle WA 98195-1580\\
	$^4$California Institute of Technology, M/C 130-33, Pasadena, CA 91125\\
         E-mail: \email{justin@physik.uzh.ch}}

\abstract{Predicting the local flux of dark matter particles is vital for dark matter direct detection experiments. To date, such predictions have been based on simulations that model the dark matter alone. Here we include the influence of the baryonic matter for the first time. We use two different approaches. Firstly, we use dark matter only simulations to estimate the expected merger history for a Milky Way mass galaxy, and then add a thin stellar disc to measure its effect. Secondly, we use three cosmological hydrodynamic simulations of Milky Way mass galaxies. In both cases, we find that a stellar/gas disc at high redshift ($z\sim1$) causes merging satellites to be preferentially dragged towards the disc plane. This results in an accreted dark matter disc that contributes $\sim 0.25 - 1$ times the non-rotating halo density at the solar position. An associated thick stellar disc forms with the dark disc and shares a similar velocity distribution. If these accreted stars can be separated from those that formed in situ, future astronomical surveys will be able to infer the properties of the dark disc from these stars. The dark disc, unlike dark matter streams, is an equilibrium structure that must exist in disc galaxies that form in a hierarchical cosmology. Its low rotation lag with respect to the Earth significantly boosts WIMP capture in the Earth and Sun, increases the likelihood of direct detection at low recoil energy, boosts the annual modulation signal, and leads to distinct variations in the flux as a function of recoil energy that allow the WIMP mass to be determined (see contribution from T. Bruch this volume).}

\FullConference{Identification of dark matter 2008\\
		 August 18-22, 2008\\
		 Stockholm, Sweden}

\begin{document}

\vspace{-3mm}
\section{Introduction}\label{sec:intro}
\vspace{-3mm}
\noindent
The case for dark matter in the Universe is based on a wide range of observational data, from galaxy rotation curves and gravitational lensing, to the Cosmic Microwave Background Radiation \cite{1996PhR...267..195J}; \cite{2006ApJ...648L.109C}. Of the many plausible dark matter candidates in extensions to the Standard Model, Weakly Interacting Massive Particles (WIMPs) stand out as well-motivated and detectable \cite{1996PhR...267..195J}, giving rise to many experiments designed to detect WIMPs in the lab. Predicting the flux of dark matter particles through the Earth is key to the success of such experiments, both to motivate detector design, and for the interpretation of any future signal \cite{1996PhR...267..195J}. 

Previous work has modelled the phase space density distribution of dark matter at solar neighbourhood using cosmological simulations that model the dark matter alone (see e.g. \cite{2008arXiv0808.2981S}, \cite{2008arXiv0809.0898S}). Here we make the first attempt to include the baryonic matter -- the stars and gas that make up the Milky Way. The Milky Way stellar disc presently dominates the mass interior to the solar radius and likely did so also in the early Universe at redshift $z=1$, when the mean merger rate in a $\Lambda$CDM cosmology peaks \cite{2008MNRAS.389.1041R}. The stellar and gas disc is important because it biases the accretion of satellites, causing them to be dragged towards the disc plane where they are torn apart by tides. The material from these accreted satellites settles into a thick disc of stars and dark matter \cite{1989AJ.....98.1554L}. In this work, we quantify the expected properties of this dark disc. Its implications for direct detection experiments and the capture of WIMPs in the Sun and Earth are presented in the contribution from T. Bruch, this volume and in \cite{2008arXiv0804.2896B}.

We use two different approaches. In the first approach (\S\ref{sec:ap1}), we use dark matter only simulations to estimate the expected merger history of a Milky Way mass galaxy, and then add a stellar disc to measure its effect. This work is presented in detail already in \cite{2008MNRAS.389.1041R}. In the second approach (\S\ref{sec:ap2}), we use cosmological hydrodynamic simulations of the Milky Way to hunt for dark discs. Both approaches are complementary in quantifying the expected properties of the dark disc. The former allows us to specify precisely the properties of the Milky Way disc at high redshift; the latter is fully self-consistent. 

\vspace{-3mm}
\section{Approach \#1: adding a stellar disc to cosmological dark matter simulations}\label{sec:ap1}
\vspace{-3mm}
\noindent
We used a cosmological dark matter only simulation already presented in \cite{2005MNRAS.364..367D} to estimate the frequency of satellite-disc encounters for a typical Milky Way galaxy; further details are given in \cite{2008MNRAS.389.1041R}. From the simulation volume, we extracted four Milky Way sized halos at a mass resolution of $m_p=5.7\times 10^5 M_\odot$. The subhalos inside each `Milky Way' and at each redshift output were identified using the algorithm in \cite{2004MNRAS.351..399G} and then traced back in time to their progenitor halos as detailed in \cite{2008MNRAS.389.1041R}. 

\begin{center}
\begin{figure}
	\setlength{\gsize}{0.26\textwidth}
	\setlength{\subfigcapskip}{-1.1\gsize}

	\hspace{0mm}
	\vspace{-3mm}
	\subfigure[\hspace{0.45\gsize}]
	{
		\label{figdda}
		\includegraphics[height=\gsize]{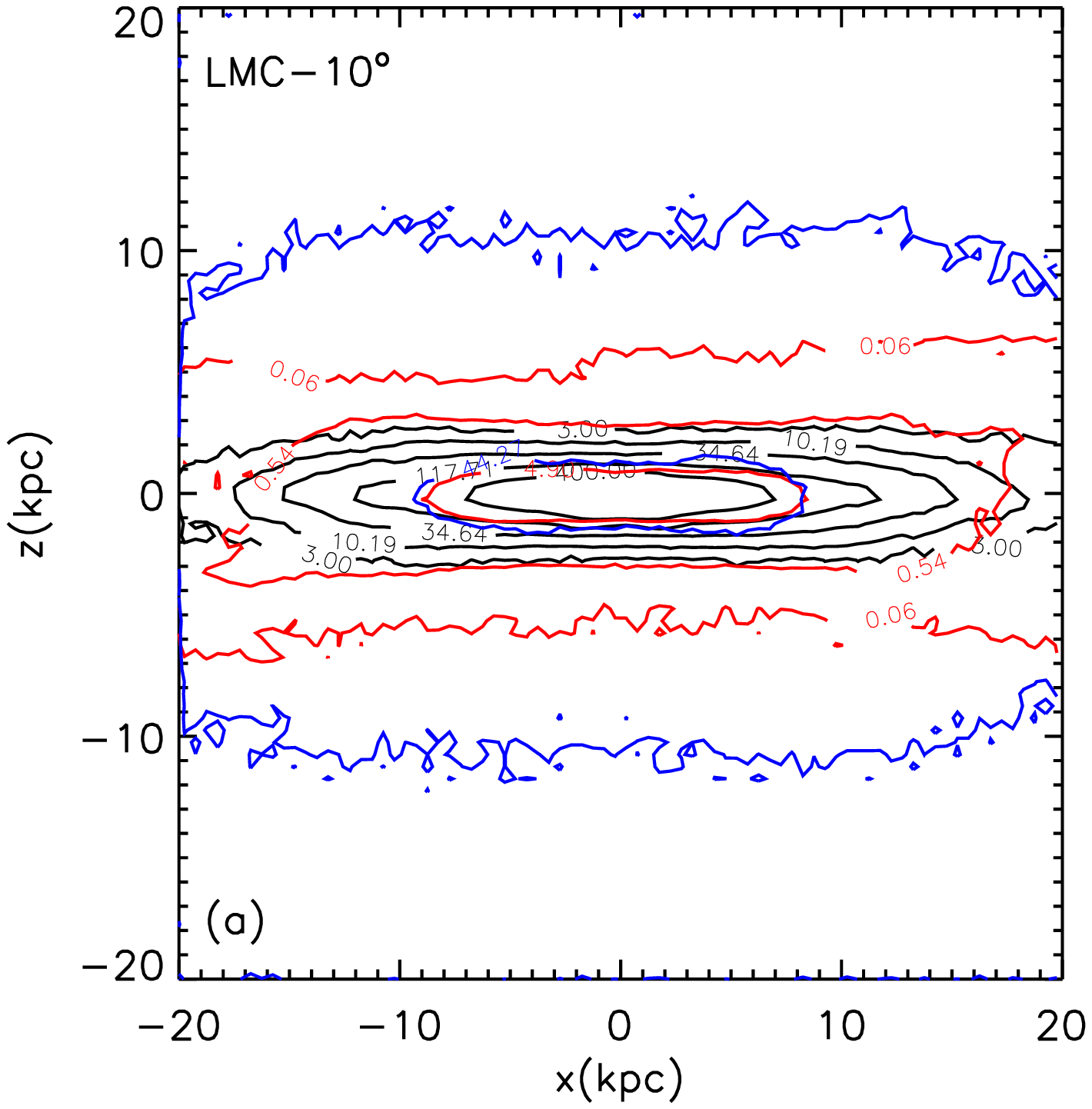}
	}
	\hspace{-0.2\gsize}
	\subfigure[\hspace{0.45\gsize}]
	{
		\label{figddb}
		\includegraphics[height=\gsize]{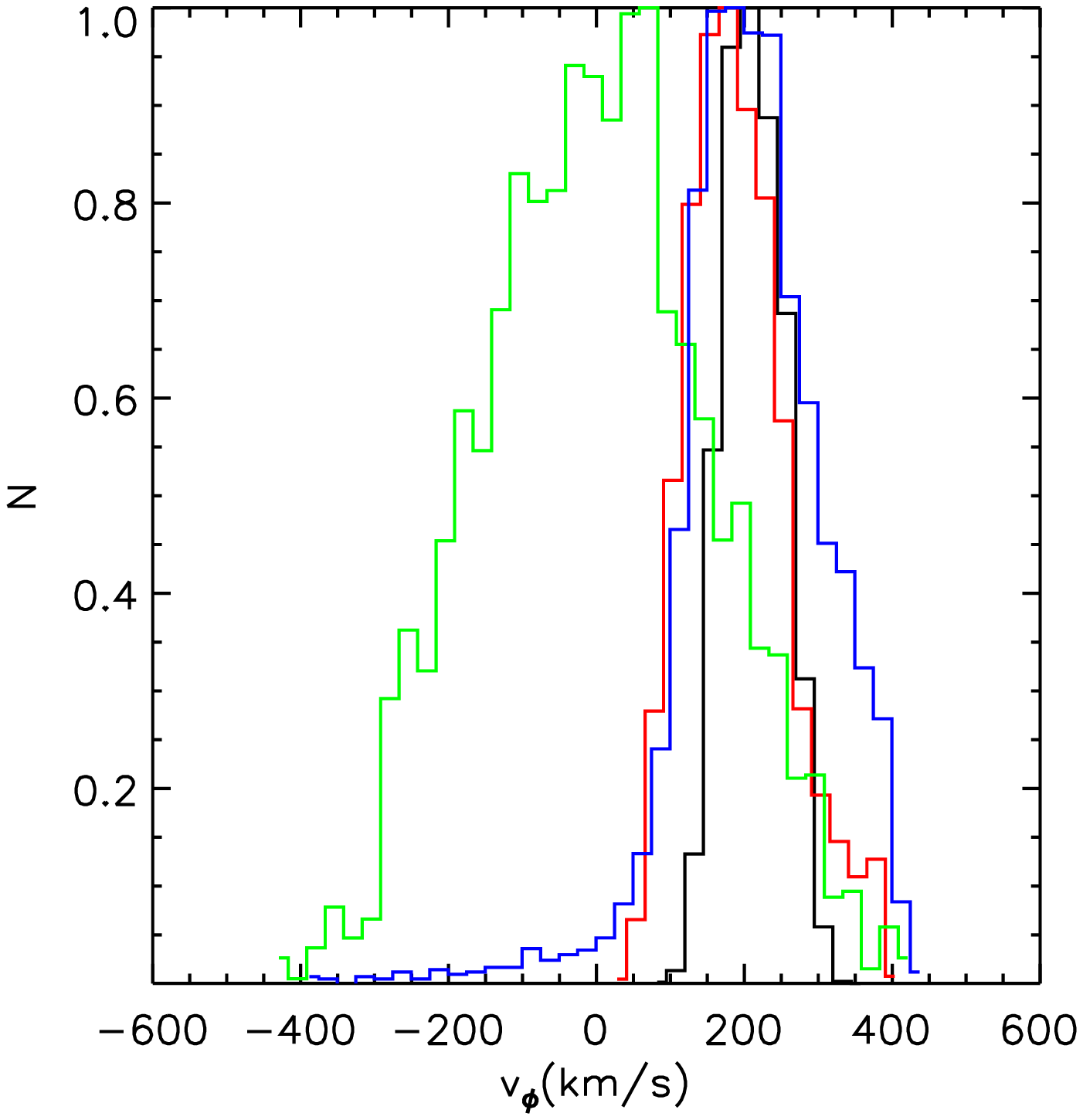}
	}
	\hspace{-0.2\gsize}
	\subfigure[\hspace{0.45\gsize}]
	{
		\label{figddc}
		\includegraphics[height=\gsize]{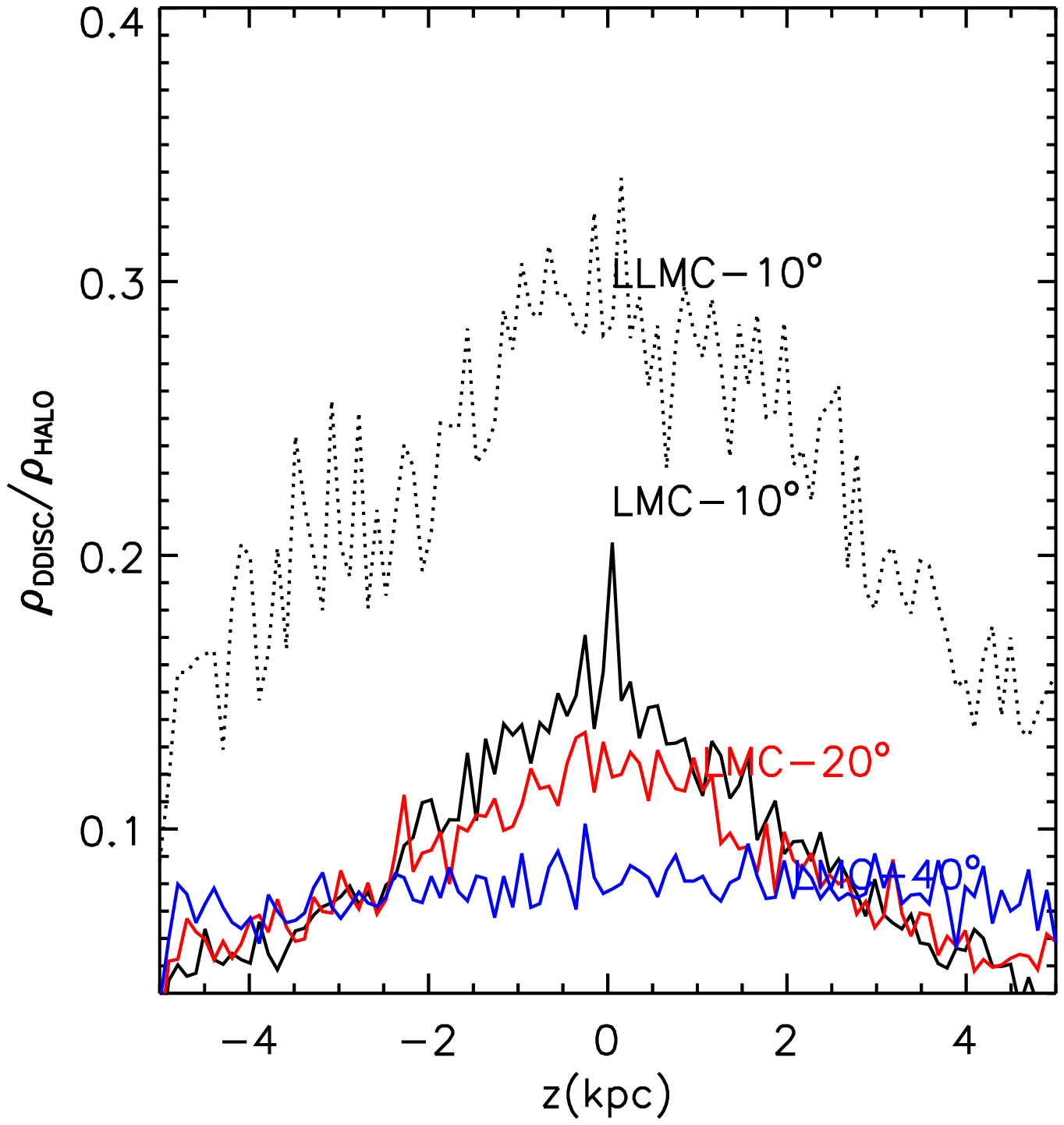}
	}
	\hspace{-0.2\gsize}
	\subfigure[\hspace{0.45\gsize}]
	{
		\label{figddd}
		\includegraphics[height=\gsize]{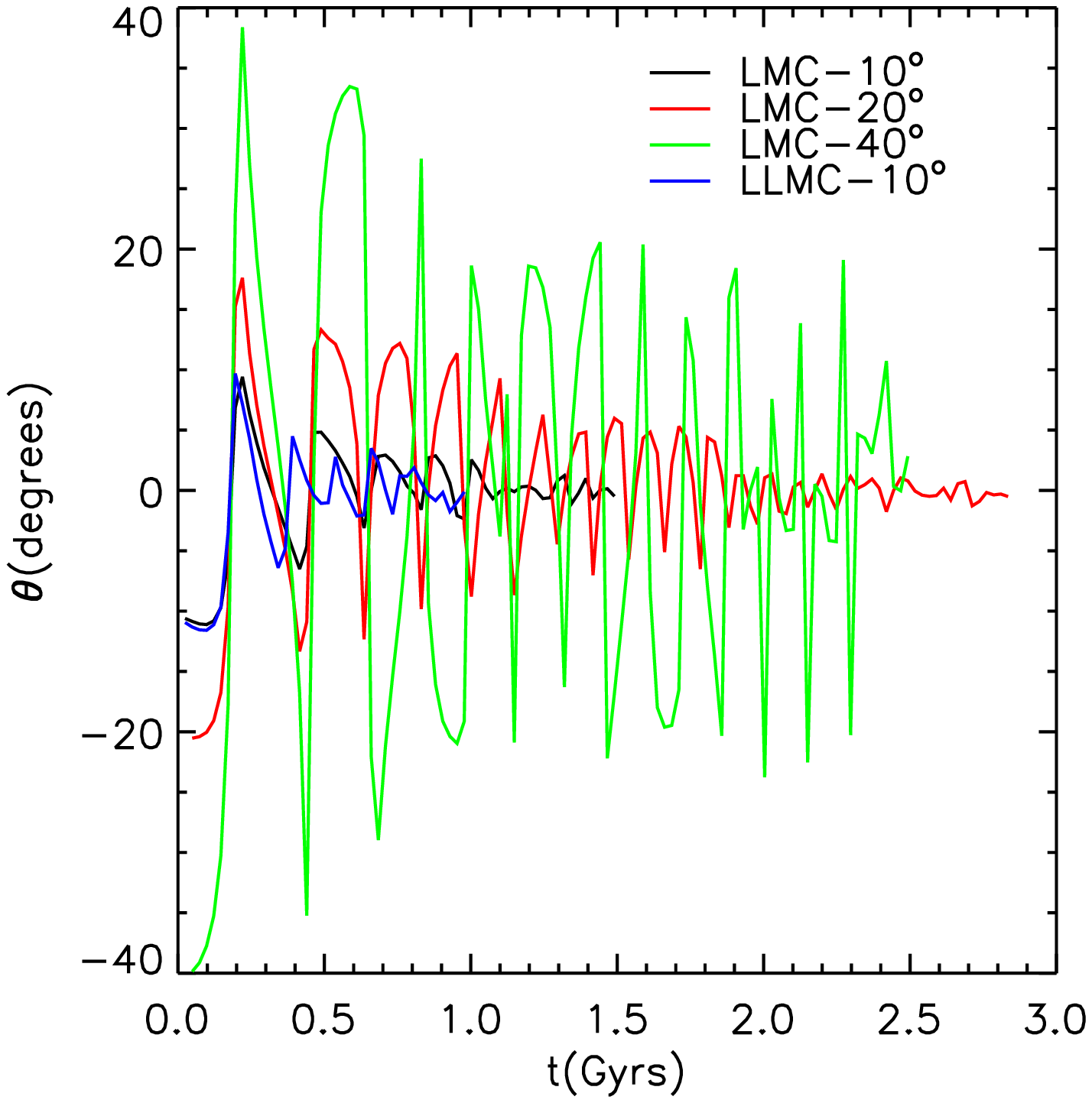}
	}
\caption{{\bf (a):} The accreted stars (red) and dark matter (blue) at the end of a simulation where the LMC satellite merged at $\theta = 10^\mathrm{o}$ to the Milky Way stellar disc; the black contours show the underlying Milky Way stellar distribution. {\bf (b):} The corresponding velocity distribution in $v_\phi$ (rotation velocity) at the solar neighbourhood; the underlying Milky Way dark matter halo is shown in green. {\bf (c):} The dark matter disc to dark matter halo density ratio $\rhodd/\rhoh$ as a function of height above the disc plane (also for $8 < R<9$\,kpc), for LMC merging at $\theta = 10^\mathrm{o}, 20^\mathrm{o}, 40^\mathrm{o}$ and LLMC at $\theta = 10^\mathrm{o}$ to the Milky Way disc plane. {\bf (d):} The satellite-disc inclination angle $\theta$ as a function of time; the lines are truncated when the satellite is fully accreted.}
\label{fig:ddresults}
\end{figure}
\vspace{-15mm}
\end{center}

From our sample of four Milky Way mass halos and assuming that mergers are isotropic (which we must do since we do not know how the disc should align with the dark matter halo), we find that a typical Milky Way sized halo will have 1($\pm 1$) subhalo merge within $\theta<20^\mathrm{o}$ of the disc plane with $\vmaxd > 80$\,km/s; $2(\pm 1)$ with $\vmaxd > 60$; and $5(\pm1)$ with $\vmaxd > 40$\,km/s. Away from the disc plane there will be twice as may mergers at the same mass \cite{2008MNRAS.389.1041R}. It is important to stress that these numbers come from the distribution of fully disrupted subhalos, not the surviving distribution that is significantly less damaging. 

We then estimated the effect of a stellar disc on these mergers by running isolated disc-merger simulations. We set up our Milky Way (MW) model (disc+halo system) by adiabatically growing a disc inside a spherical halo, as detailed in \cite{2008MNRAS.389.1041R}. We chose three models for our satellite, but present just two here: LMC with $\vmaxd=60$\,km/s, and LLMC with $\vmaxd=80$\,km/s; these were set up as scaled versions of our MW model. We chose a wide range of initial inclination angles to the disc from $\theta = 10-60^\mathrm{o}$, one retrograde orbit, and range of pericentres and apocentres. The simulations were evolved using the collisionless tree-code, PkdGRAV \cite{2001PhDT........21S}. The final evolved systems were mass and momentum centred using the `shrinking sphere' method described in \cite{2006MNRAS.tmp..153R}, and rotated into their moment of inertia eigenframe with the $z$ axis perpendicular to the disc. 

The results are shown in Figure \ref{fig:ddresults}. The left panel shows the accreted stars (red) and dark matter (blue) at the end of a simulation where the LMC satellite merged at $\theta = 10^\mathrm{o}$ to the disc. Both the stars and the dark matter have settled into accreted discs. The middle panel shows the corresponding velocity distribution in $v_\phi$ (rotation velocity) at the solar neighbourhood (a cylinder $8<R<9$\,kpc, $|z|<0.35$\,kpc). The underlying dark matter halo is shown in green and is not rotating; the accreted stars and dark matter (red and blue) have kinematics similar to that of the underlying stellar disc (black). The right panel shows the dark matter disc to dark matter halo density ratio $\rhodd/\rhoh$ as a function of height above the disc plane for selected merger simulations, as marked. As the satellite impact angle $\theta$ is increased, the satellite contributes less material to a dark disc. For $\theta = 40^\mathrm{o}$, the density at the solar neighbourhood is nearly flat with $z$ and less than a tenth of the underlying halo density; there is correspondingly less rotation in this simulation. Summing over the expected number and mass of mergers, we find that the dark disc contributes $\sim 0.25 - 1$ times the non-rotating halo density at the solar position \cite{2008MNRAS.389.1041R}. It is important to stress that all satellites regardless of their initial inclination have some accreted material that is focused into the disc plane (see Figure \ref{fig:ddresults}), right panel. As such, we expect that the accreted dark and stellar discs will comprise several accreted satellites; the most massive low-inclination mergers being the most important contributors.

The accreted stellar disc shares similar kinematics to the dark disc. Depending on assumptions about the mass to light ratio of accreted satellites, these accreted stars can make up $\sim 10 - 50$\% of the Milky Way stellar thick disc \cite{2008MNRAS.389.1041R}. (The lower end of this range is more likely given the observed properties of satellite galaxies in the Universe today.) If future surveys of our Galaxy can disentangle accreted stars in the Milky Way thick disc from those that formed in-situ, then we will be able to infer the velocity distribution function of the dark disc from these stars. 

\vspace{-3mm}
\section{Approach \#2: cosmological hydrodynamic simulations}\label{sec:ap2}
\vspace{-3mm}
\noindent
We use three cosmological hydrodynamic simulations of Milky Way mass galaxies, two of which (\MWone,\MWthree) have already been presented in \cite{2008arXiv0801.3845M} and \cite{2008ASPC..396..453G}. All three were run with the GASOLINE code \cite{2004NewA....9..137W} using the "blastwave feedback" described in \cite{2006MNRAS.373.1074S}. \MWone\ had cosmological parameters $(\Omega_{\rm m},\Omega_{\Lambda},\sigma_8,h)=(0.3,0.7,0.9,0.7)$; \MWtwo\ and \MWthree\ used $(\Omega_{\rm m},\Omega_{\Lambda},\sigma_8,h)=(0.24,0.76,0.77,0.73)$. At redshift $z=0$ the typical particle masses for dark matter stars and gas were: $(M_{dm},M_*,M_{gas}) = (7.6,0.2,0.3)\times 10^5\,$M$_\odot$, with associated force softening: $(\epsilon_{dm},\epsilon_*,\epsilon_{gas}) = (0.3,0.3,0.3)\,$kpc. The analysis was performed as in \S\ref{sec:ap1}.

\begin{center}
\begin{figure}
	\setlength{\gsize}{0.24\textwidth}
	\setlength{\subfigcapskip}{-1.1\gsize}

	\hspace{0mm}
	\vspace{-3mm}
	\subfigure[\MWone\hspace{0.28\gsize}]
	{
		\label{fig2a}
		\includegraphics[height=\gsize]{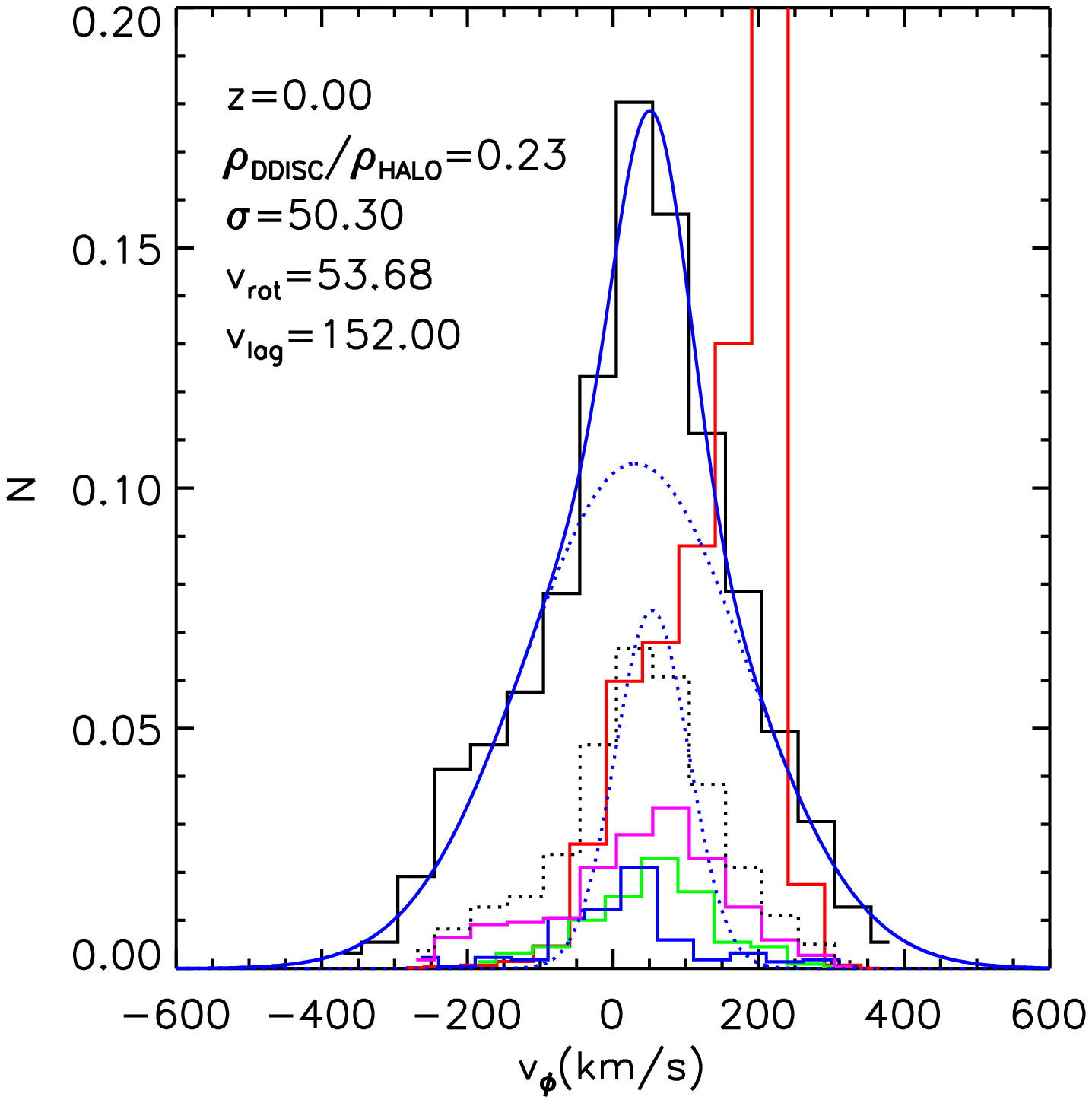}
	}
	\hspace{-0.2\gsize}
	\subfigure[\MWtwo\hspace{0.28\gsize}]
	{
		\label{fig2b}
		\includegraphics[height=\gsize]{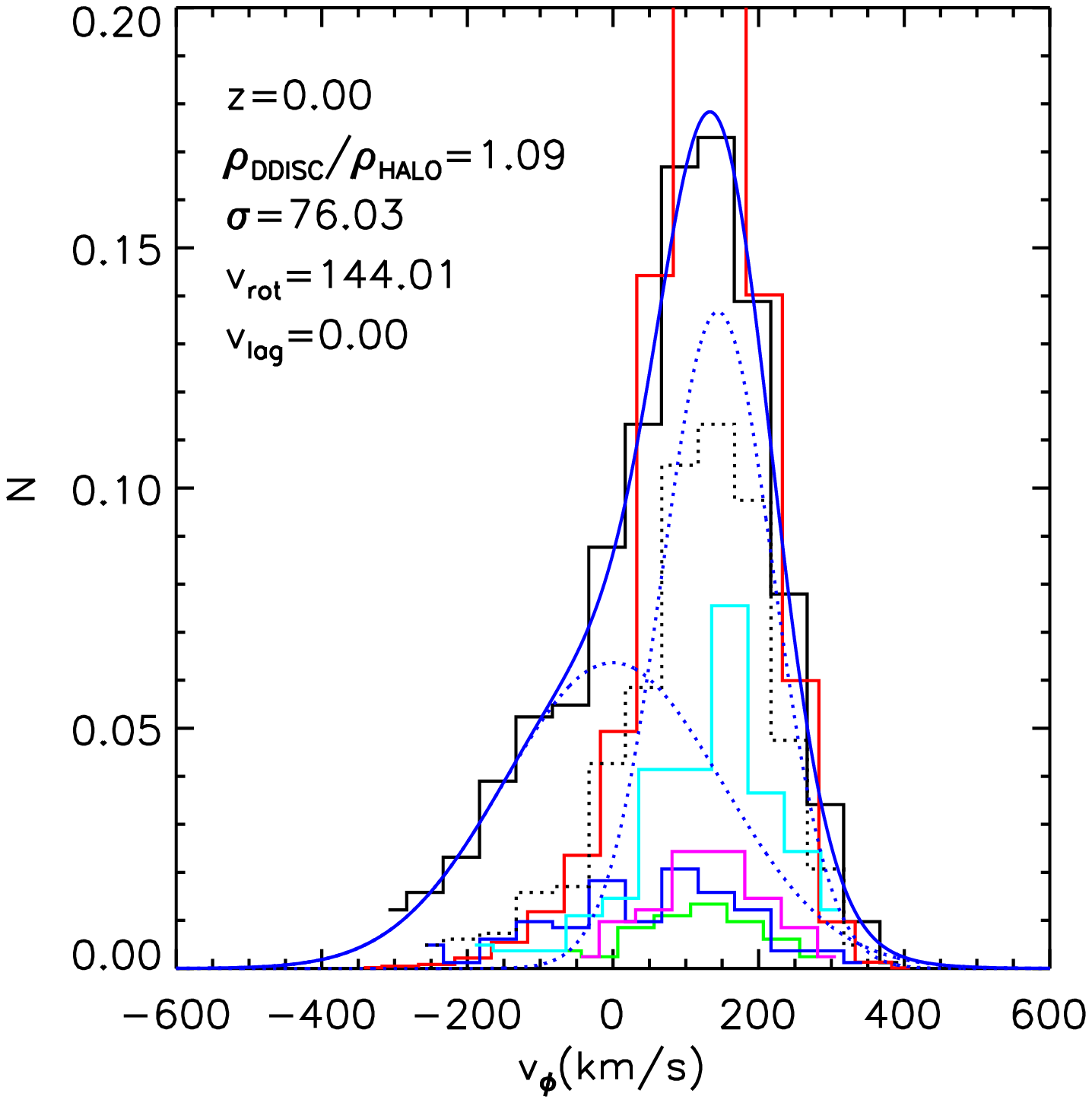}
	}
	\hspace{-0.2\gsize}
	\subfigure[\MWthree\hspace{0.28\gsize}]
	{
		\label{fig2c}
		\includegraphics[height=\gsize]{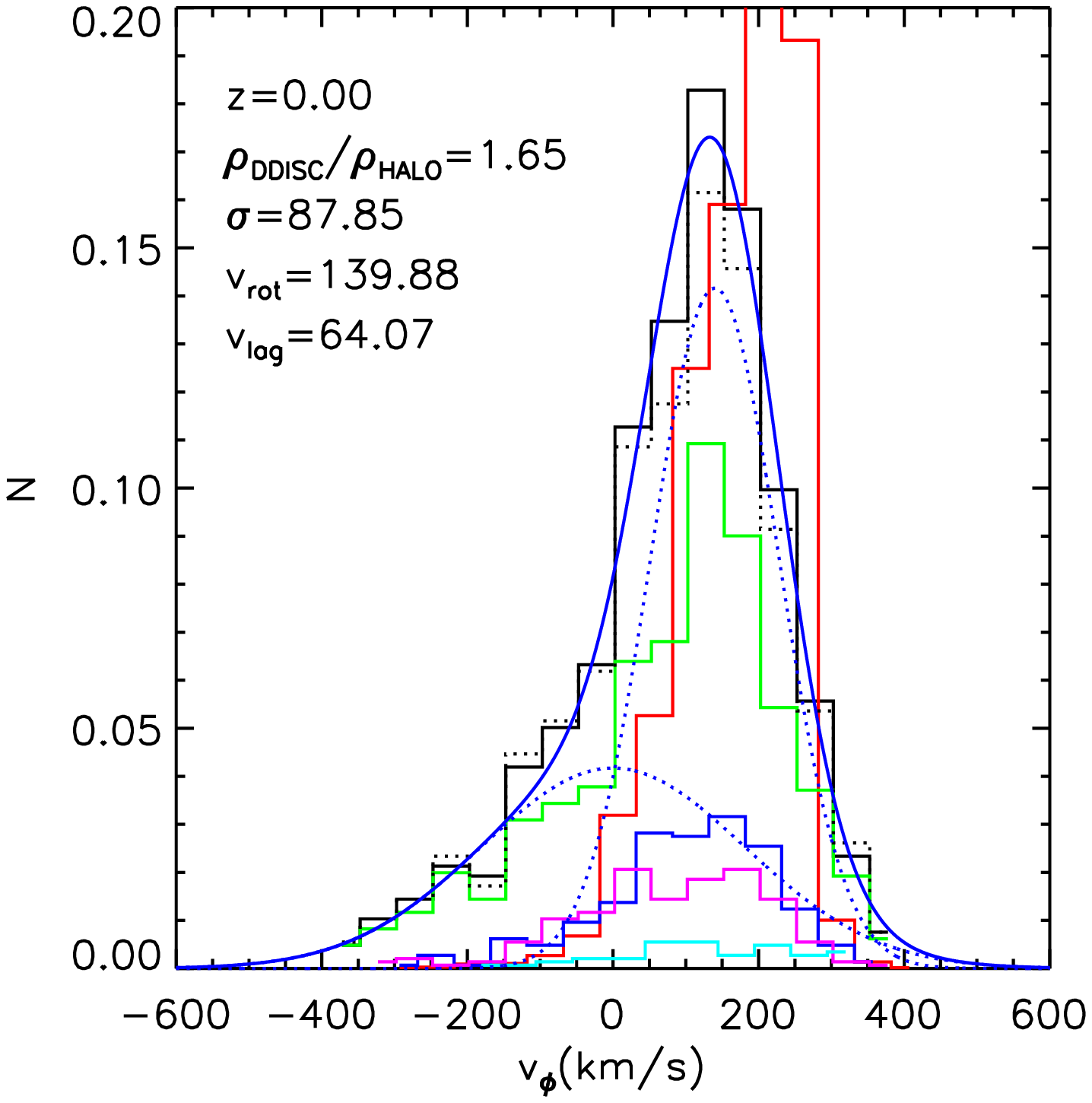}
	}
	\hspace{-0.2\gsize}
	\subfigure[\MWthreedark\hspace{0.15\gsize}]
	{
		\label{fig2d}
		\includegraphics[height=\gsize]{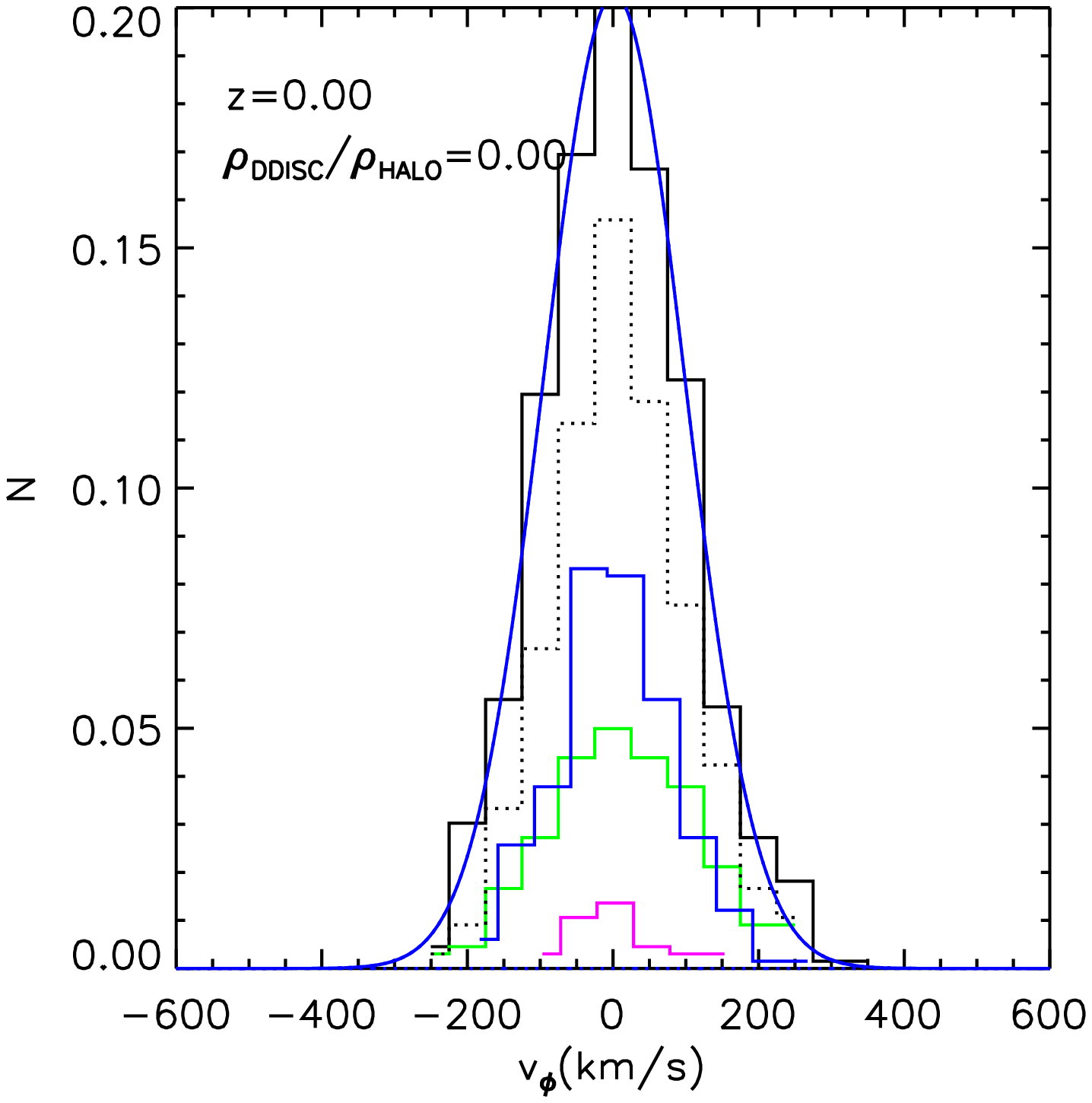}
	}\\
	
	\subfigure[\MWone\hspace{0.28\gsize}]
	{
		\label{fig2e}
		\includegraphics[height=\gsize]{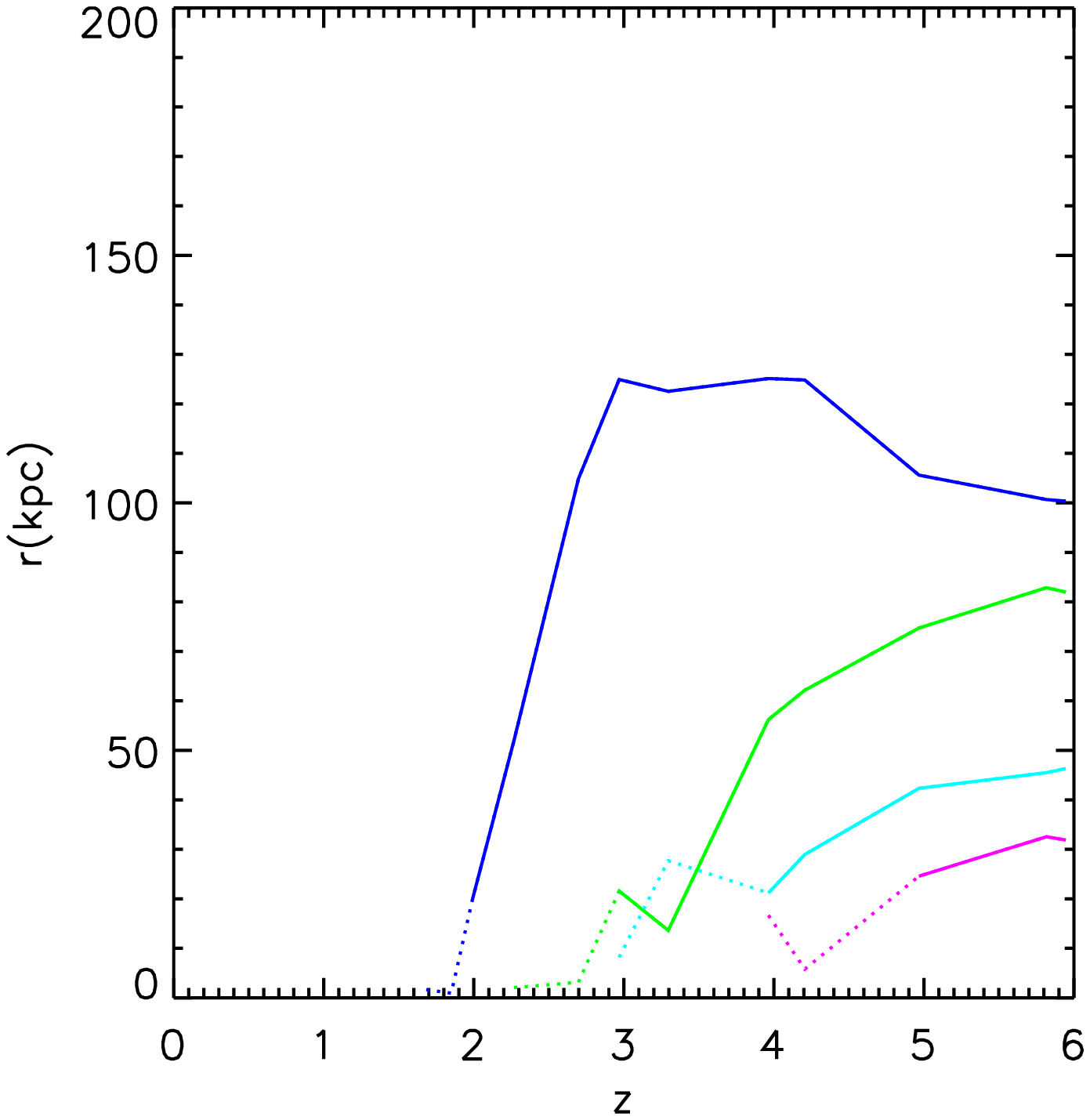}
	}
	\hspace{-0.2\gsize}
	\subfigure[\MWtwo\hspace{0.28\gsize}]
	{
		\label{fig2f}
		\includegraphics[height=\gsize]{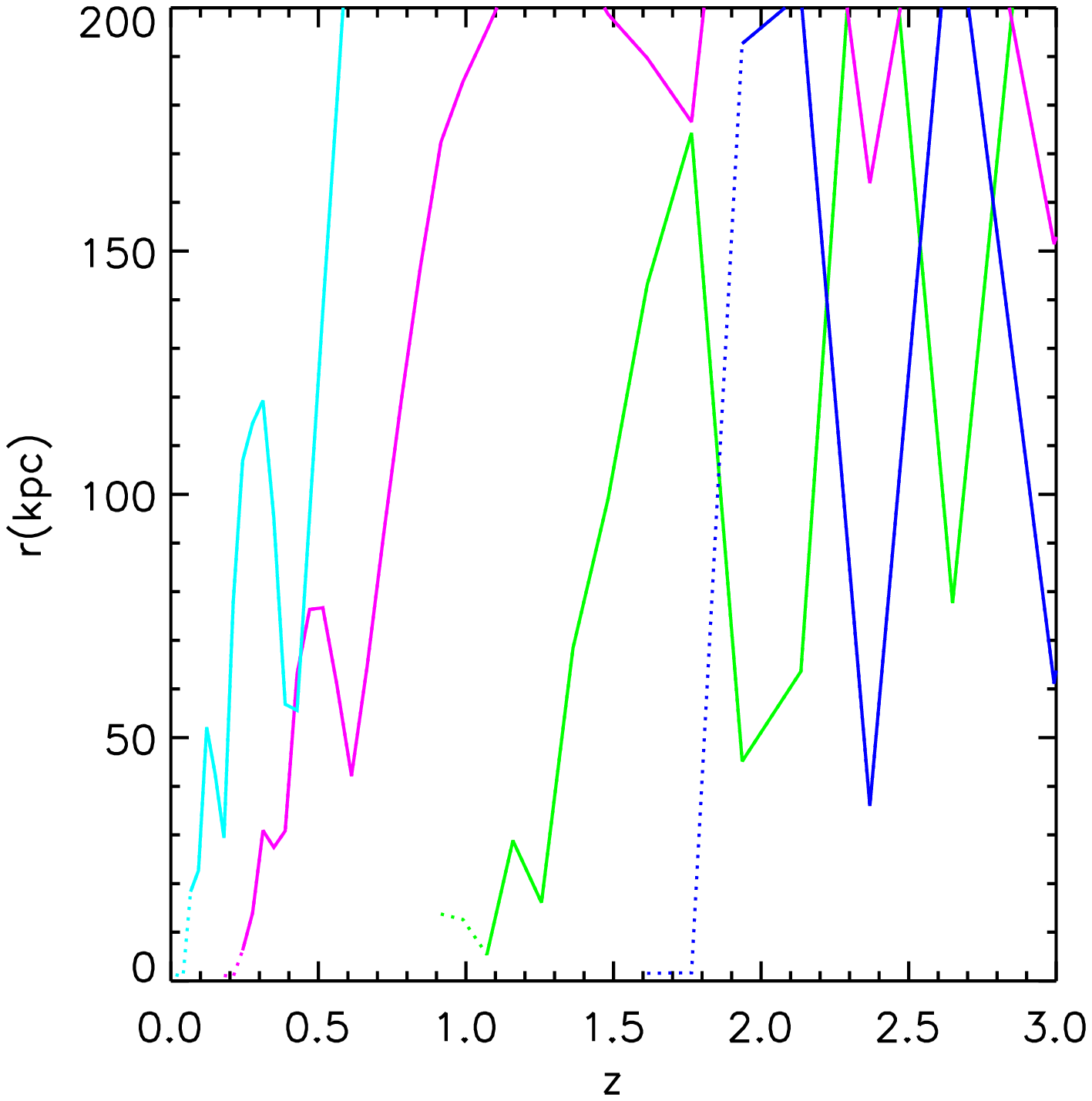}
	}
	\hspace{-0.2\gsize}
	\subfigure[\MWthree\hspace{0.28\gsize}]
	{
		\label{fig2g}
		\includegraphics[height=\gsize]{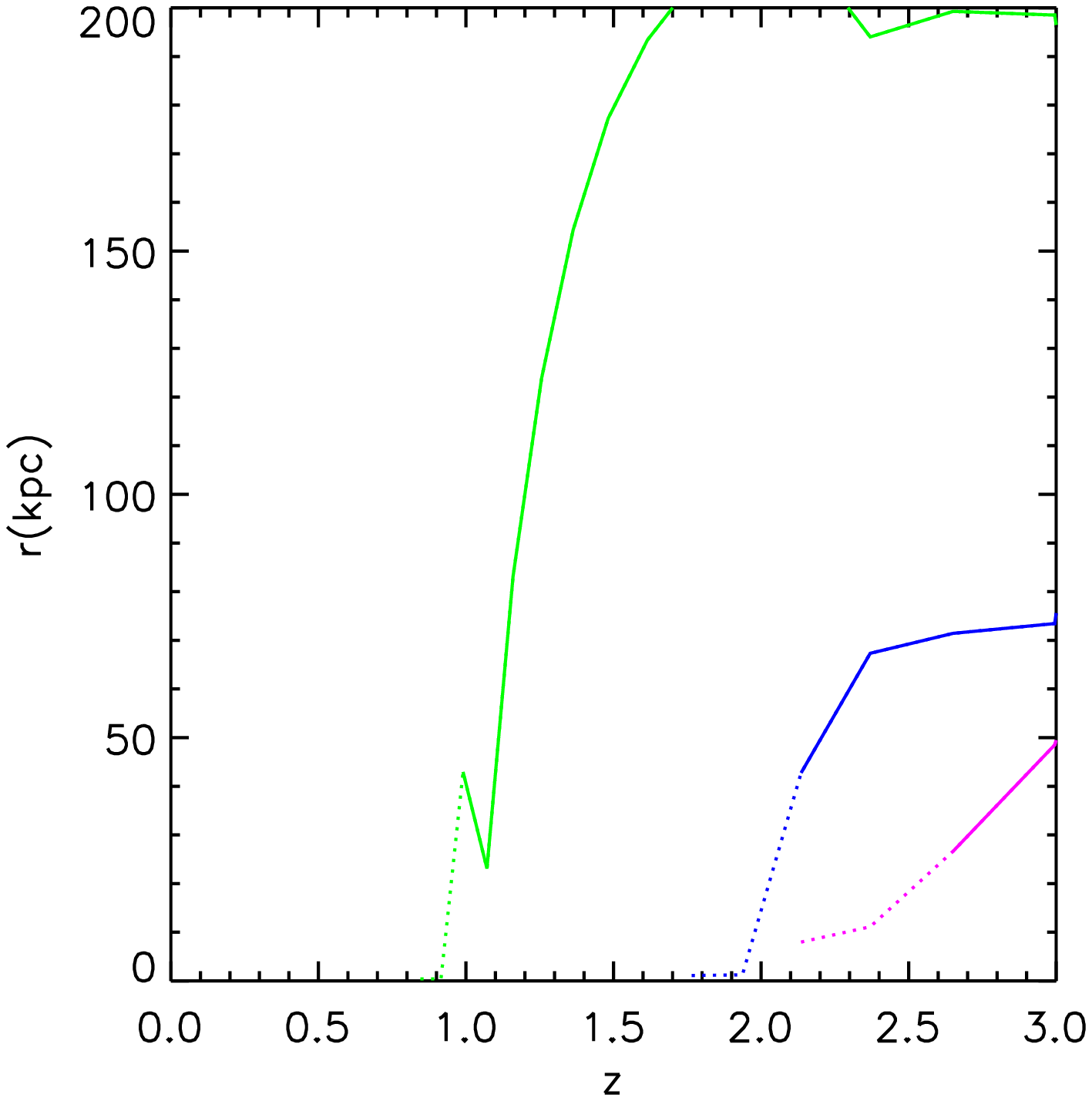}
	}
	\hspace{-0.2\gsize}
	\subfigure[\MWthreedark\hspace{0.15\gsize}]
	{
		\label{fig2h}
		\includegraphics[height=\gsize]{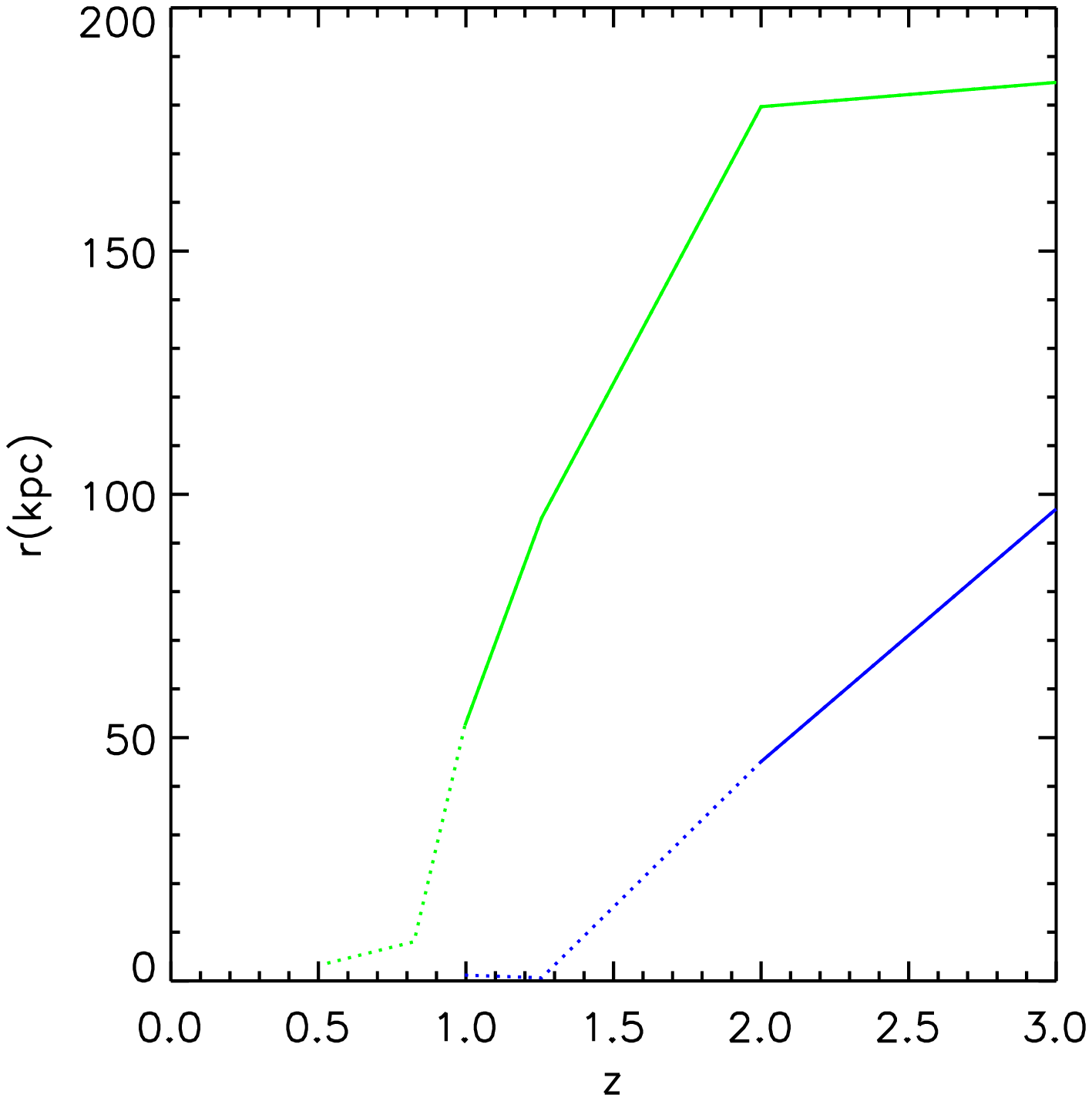}
	}
\vspace{-6mm}
\caption{{\bf (a-c)} The distribution of rotational velocities at the solar neighbourhood for three simulated Milky Way mass galaxies \MWone, \MWtwo\ and \MWthree. The lines show the dark matter (black), stars (red), a double Gaussian fit to the dark matter (blue; blue dotted), the material accreted from the four most massive disrupted satellites (green, blue, magenta, cyan), and the sum of all material accreted from these satellites (black dotted). The best fit double Gaussian parameters are marked in the top left, along with the redshift, $z$. {\bf (d)} As (a-c), but for the galaxy \MWthree\ simulated with {\it dark matter alone}. {\bf (e-h)} The decay in radius $r$ as a function of redshift $z$ of the four most massive disrupting satellites in \MWone, \MWtwo\, \MWthree\ and \MWthreedark. Where less than four lines are shown, these satellites accreted at redshift $z>3$. The dotted sections show the evolution of the most bound satellite particle after the satellite has disrupted.}
\label{fig:cosmo}
\vspace{-3mm}
\end{figure}
\vspace{-15mm}
\end{center}

Figure \ref{fig:cosmo} shows the distribution of rotational velocities at the solar neighbourhood (top panels), and the orbital decay of the four most massive satellites (bottom panels) for \MWone, \MWtwo\ and \MWthree. The right most panels (d) and (h) show the results for the galaxy \MWthree\ simulated {\it without} any gas or stars -- \MWthreedark.

Notice that, with the exception of \MWthreedark, in all cases the dark matter requires a double Gaussian fit to its local $v_\phi$ velocity distribution. Once the baryons (the stars and gas) are included in the simulation, there is a local dark matter disc that lags the rotation of the thin stellar disc by $\sim 50-150$\,km/s. The mass and rotation speed of the dark disc increase for the simulations that have more late mergers. \MWone\ has no significant mergers after redshift $z=2$ and has a less significant dark disc, with rotation lag $\sim150$\,km/s and dark disc to non-rotating dark halo density ratio $\rhodd/\rhoh = 0.23$ (obtained from the double Gaussian fit). \MWtwo\ and \MWthree\ both have extreme dark discs with $\rhodd/\rhoh > 1$ and rotation lag $\simlt 60$\,km/s; they both have massive mergers at redshift $z<1$. 

\vspace{-3mm}
\section{Conclusions}\label{sec:conclusions}
\vspace{-3mm}
\noindent
Low inclination massive satellite mergers are expected in a $\Lambda$CDM cosmology. These lead to the formation of thick accreted dark and stellar discs. We used two different approaches to estimate the importance of the dark disc. Firstly, we used dark matter only simulations to estimate the expected merger history for a Milky Way mass galaxy, and then added a thin stellar disc to measure its effect. Secondly, we used three cosmological hydrodynamic simulations of Milky Way mass galaxies. In both cases, we found that a typical Milky Way mass galaxy will have a dark disc that contributes $\sim 0.25 - 1$ times the non-rotating halo density at the solar position. The dark disc has important implications for the direct detection of dark matter \cite{2008arXiv0804.2896B} and for the capture of WIMPs in the Sun and Earth (see contribution from T. Bruch, this volume). 

\bibliographystyle{aa}
\bibliography{refs}

\end{document}